\documentclass[aps,pra,preprint,groupedaddress,showpacs,showkeys]{revtex4}
\usepackage{amssymb,bm}
\usepackage{graphicx}

\begin{document}

\title{ High-energy expansion of Coulomb corrections to the
        $e^+e^-$ photoproduction cross section}

\author{R.N. Lee}\email{R.N.Lee@inp.nsk.su}
\author{A. I. Milstein}\email{A.I.Milstein@inp.nsk.su}
\author{V.M. Strakhovenko}\email{V.M.Strakhovenko@inp.nsk.su}

\affiliation{Budker Institute of Nuclear Physics, 630090 Novosibirsk, Russia}

\date{\today}

\begin{abstract}
First correction to the high-energy asymptotics of the total $e^+e^-$
photoproduction cross section in the electric field of a heavy atom is derived
with the exact account of this field. The consideration is based on the use of
the quasiclassical electron Green function in an external electric field.  The
next-to-leading correction to the cross section is discussed. The influence of
screening on the Coulomb corrections is examined in the leading approximation.
It turns out that the high-energy asymptotics of the corresponding correction
is independent of the photon energy. In the region where both produced
particles are relativistic, the corrections to the high-energy asymptotics of
the electron (positron) spectrum are derived. Our results for the total cross
section are in good agreement with experimental data for photon energies down
to a few $MeV$. In addition, the corrections to the bremsstrahlung spectrum are
obtained from the corresponding results for pair production.
\end{abstract}

\pacs{32.80.-t, 12.20.Ds}

\keywords{ $e^+e^-$ photoproduction, bremsstrahlung, Coulomb corrections,
screening}

\maketitle



\section{Introduction}

Knowledge of the photoabsorption cross sections is very important in various
applications, see, e.g., \citep{Hubbell2000}. The relevant processes are the
atomic photoeffect, nuclear photoabsorption, incoherent and coherent photon
scattering and $e^+e^-$ pair production. In the coherent processes, by
definition, there is no excitation or ionization of an atom. The high-accuracy
estimation of the corresponding cross sections is required. They have different
dependence on the photon energy $\omega$. At $\omega\gtrsim 10 MeV$, the cross
section  of $e^+e^-$ pair production becomes dominant \citep{HGO1980}. The
coherent contribution $\sigma_{coh}$ to the pair production cross section is
roughly $Z$ times larger than the incoherent one ($Z$ is the atomic number),
thereby being the most important for heavy atoms. Just the coherent pair
production is considered below.

The theoretical and experimental investigation of the coherent pair production
has a long history, see \citep{HGO1980}. In the Born approximation, the cross
section $\sigma_B$ is known for arbitrary photon energy
\citep{BH1934,Racah1934}. The account of the effect of screening is
straightforward in this approximation and can be easily performed if the atomic
form factor is known \citep{JLS1950}. For heavy atoms it is necessary to take
into account the Coulomb corrections $\sigma_C$,
\begin{equation}\label{eq:sigmacoh}
\sigma_{coh}=\sigma_B+\sigma_C\,.
\end{equation}

These corrections are higher order terms of the perturbation theory with
respect to the atomic field. The magnitude of $\sigma_C$ depends on $\omega$
and the parameter $Z\alpha$ ($\alpha=1/137$ is the fine-structure constant).
The formal expression for $\sigma_C$, exact in $Z\alpha$ and $\omega$, was
derived by \citet{Overbo1968}. This expression has a very complicated form
causing severe difficulties in computations. The difficulties grow as $\omega$
increases, so that numerical results in \citep{Overbo1968} were obtained only
for $\omega<5 MeV$.

In the high-energy region $\omega\gg m$ ($m$ is the electron mass,
$\hbar=c=1$), the consideration is greatly simplified. As a result, a rather
simple form was obtained in \citep{BM1954,DBM1954} for the Coulomb corrections
in the leading approximation with respect to $m/\omega$. However, the
theoretical description of the Coulomb corrections at intermediate photon
energies ($5\div100 MeV$) has not been completed. At present, all estimates of
$\sigma_C$ in this region are based on the "bridging" expression derived by
\citet{Overbo1977}. This expression is actually an extrapolation of the results
obtained for $\omega<5MeV$. It is based on some assumptions on the form of the
asymptotic expansion of $\sigma_C$ at high photon energy. It is commonly
believed that the "bridging" expression has an accuracy providing the maximum
error in $\sigma_{coh}$ of the order of a few tens of percent.

Here we develop a description of $e^+e^-$-pair production at intermediate
photon energies by deriving the next-to-leading term of the high-energy
expansion of $\sigma_C$. First we consider a pure Coulomb field and represent
$\sigma_C$ in the form
\begin{equation}\label{eq:expansion}
\sigma_C=\sigma_C^{(0)}+\sigma_C^{(1)}+\sigma_C^{(2)}+\ldots
\end{equation}
The term $\sigma_C^{(n)}$ has the form $(m/\omega)^n S^{(n)}(\ln \omega/m)$,
where $S^{(n)}(x)$ is some polynomial. The $\omega$-independent term
$\sigma_C^{(0)}$ corresponds to the result of \citet{DBM1954}. In the present
paper we derive the term $\sigma_C^{(1)}$. It turns out that $S^{(1)}$ is
$\omega$-independent in contrast to a second-degree polynomial suggested by
\citet{Overbo1977}. We present an ansatz for $\sigma_C^{(2)}$, which provides a
good agreement with available experimental data for $\omega>5 MeV$.

The high-energy expansion of the Coulomb corrections to the spectrum has the
form similar to (\ref{eq:expansion}). In the region $\varepsilon_{\pm}\gg m$,
we derive the term $d\sigma_C^{(1)}/dx$, where $\varepsilon_-$ and
$\varepsilon_+$ are the electron and positron energy, respectively,
$x=\varepsilon_{-}/\omega$. The term $d\sigma_C^{(1)}/dx$ may turn important,
e.g., for description of the development of electromagnetic showers in a
medium. The correction found is antisymmetric with respect to the permutation
$\varepsilon_{+}\leftrightarrow\varepsilon_{-}$ and does not contribute to the
total cross section. In fact,  $\sigma_C^{(1)}$ originates from two energy
regions $\varepsilon_{+}\sim m$ and $\varepsilon_{-}\sim m$, where the spectrum
is not known. However, our result for $\sigma_C^{(1)}$ allows us to claim that
the spectrum in these regions differs drastically from the result obtained by
\citet{DBM1954} for $\varepsilon_{\pm}\gg m$, if the latter is formally applied
at $\varepsilon_-\sim m$ or $\varepsilon_+\sim m$.

The effect of screening on $\sigma_C$ at $\omega\gg m$ is considered
quantitatively. In the leading approximation, we find the corresponding
correction $\sigma_C^{(scr)}$, which is $\omega$-independent similar to
$\sigma_C^{(0)}$. So, for the atomic field, $\sigma_C^{(scr)}$ should be added
to the right-hand side of Eq. (\ref{eq:expansion}). The screening correction to
the spectrum is also obtained.

In this paper we present the explicit calculations of the corrections, which
have been given without derivation in our recent work \cite{LMS2003} and used
for the detailed comparison of theory with experimental data.

\section{General discussion}

The cross section of $e^+e^-$ pair production by a photon in an
external field reads
\begin{equation}\label{eq:cs}
d\sigma_{coh}=\frac{\alpha}{(2\pi)^4\omega}\,d\bm{p}\,d\bm{q}\,\delta
(\omega - \varepsilon_+ -\varepsilon_-)|M|^{2}\,,
\end{equation}
where $\varepsilon_-=\varepsilon_{ p}=\sqrt{\bm p^2+m^2}$,
$\varepsilon_+=\varepsilon_{ q}$, and $\bm p$, $\bm q$ are the electron and
positron momenta, respectively. The matrix element $M$ has the form
\begin{equation}
M\,=\,\int d\bm r \,\bar \psi_{\bm p }^{(+)}(\bm r )\,\hat
{e}\,\psi _{\bm q}^{(-)}(\bm r )\exp{(i\bm k\bm r )}\,\,.
\end{equation}
Here $ \psi_{\bm p}^{(+)}$ and $\psi_{\bm q}^{(-)}$ are positive-energy and
negative-energy solutions of the Dirac equation in the external field,
$e_{\mu}$ is the photon polarization 4-vector,   $\bm k$ is the photon
momentum, $\hat{e}=e_{\mu}\gamma^\mu$, $\gamma^\mu$ are the Dirac matrices. It
is convenient to study various processes in external fields using the Green
function $G(\bm{r}_2,\bm{r}_1|\varepsilon)$ of the Dirac equation in this
field. This Green function can be represented in the form
\begin{eqnarray}\label{eq:FG}
G(\bm r_2 ,\bm r_1|\,\varepsilon)\,&=&\,\sum_{n}
\frac{\psi_{n}^{(+)}(\bm r_2)\bar\psi_{n}^{(+)}(\bm r_1 )}
{\varepsilon -\varepsilon_{n}\, + i0} \nonumber\\
&&+ \int \frac{d\bm{p}}{(2\pi)^{3}}\,\left[\,\frac{\psi_{\bm p}^{(+)}(\bm r_2 )
\bar\psi_{\bm p}^{(+)}(\bm r_1 )}{\varepsilon -\varepsilon_{p}\,+ i0}\,+
\,\frac{\psi_{\bm p}^{(-)}(\bm r_2 )\bar\psi_{\bm p}^{(-)}(\bm r_1 )}
{\varepsilon +\varepsilon_{p}\,- i0}\,\right] \,\,,
\end{eqnarray}
where $\psi_{n}^{(+)}$ is the discrete-spectrum wave function,
$\varepsilon_{n}$ is the corresponding binding energy. The regularization of
denominators in (\ref{eq:FG}) corresponds to the Feynman rule. From
(\ref{eq:FG}),
\begin{eqnarray}\label{eq:FG1}
\int d\Omega_{\bm{q}}\ \psi_{\bm q}^{(-)}(\bm r_2 ) \bar\psi_{\bm q}^{(-)}(\bm
r_1 )&=& -i \frac{(2\pi)^2}{q\varepsilon_{ q}}
 \delta G\,(\bm r_2 ,\bm
r_1|-\varepsilon_{\bm q})\,,\nonumber\\
\int d\Omega_{\bm{p}}\ \psi_{\bm p}^{(+)}(\bm r_1 ) \bar\psi_{\bm
p}^{(+)}(\bm r_2 )&=& i \frac{(2\pi)^2}{p\varepsilon_{\bm p}}
 \delta G\,(\bm r_1 ,\bm r_2|\varepsilon_{ p})\,,
\end{eqnarray}
where $\Omega_{\bm{p}}$ is the solid angle of $\bm p$, and $\delta
G=G-\tilde{G}$. The function $\tilde{G}$ is obtained from
(\ref{eq:FG}) by the replacement $i0\leftrightarrow -i0$.

Taking the integrals over $\Omega_{\bm{p}}$ and $\Omega_{\bm{q}}$
in (\ref{eq:cs}), we obtain the electron  spectrum, which is
 the cross section differential with respect to the  electron energy
 $\varepsilon_-$ .  Using relations
(\ref{eq:FG1}), we express this spectrum  via the Green functions:
\begin{eqnarray}\label{eq:se}
 \frac{d\sigma_{coh}}{d\varepsilon_-}=
 \frac{ \alpha}{\omega}
\int\!\!\!\!\int d\bm r_1\, d\bm r_2\, \mbox{e}^{-i\bm{kr}}\,Sp\,\left\{
 \delta G(\bm r_2,\bm r_1|\varepsilon_-)\,\hat{e}\,
 \delta G(\bm r_1,\bm r_2|-\varepsilon_+)\,\hat{e}\right\}\,,
\end{eqnarray}
where $\bm r=\bm r_2-\bm r_1$ and $\varepsilon_+=\omega-\varepsilon_-$ is the
positron energy. Since the spectrum is independent of the photon polarization,
here and below we assume $\bm e^*=\bm e$ (linear polarization).

Due to the optical theorem, the process of pair production is related to the
process of Delbr\"uck scattering (coherent scattering of a photon in the
electric field of an atom via virtual electron-positron pairs). At zero
scattering angle, the amplitude $M_D$ of Delbr\"uck scattering reads
\begin{eqnarray}\label{eq:Md}
M_D=2i\alpha \int\!\! d\varepsilon \int\!\!\!\!\int d\bm r_1\, d\bm r_2\,
\mbox{e}^{-i\bm{kr}}\,Sp\,\left\{
 G(\bm r_2,\bm r_1|\varepsilon)\,\hat{e}\,
 G(\bm r_1,\bm r_2|\varepsilon-\omega)\,\hat{e}\right\}.
\end{eqnarray}
It is necessary to subtract, from the integrand in (\ref{eq:Md}), the value of
this integrand at zero external field ($Z\alpha=0$). Below, such a subtraction
is assumed to be made.

It follows from Eqs. (\ref{eq:se}),(\ref{eq:Md}) and the analytical properties
of the Green function that
\begin{equation}\label{eq:Opt}
\frac1\omega\mbox{Im}\,M_D=\sigma_{coh}+\sigma_{bf}\,.
\end{equation}
Here
\begin{eqnarray}\label{eq:bf}
&& \sigma_{bf}=
 -\frac{ 2i\pi\alpha}{\omega}
\int\!\!\!\!\int d\bm r_1\, d\bm r_2\, \mbox{e}^{-i\bm{kr}}\sum_n\, Sp\left\{
\rho_n(\bm r_2,\bm r_1)\,\hat{e}\,
 \delta G(\bm r_1,\bm r_2|\varepsilon_n-\omega)\,\hat{e}\right\},\nonumber\\
&&\rho_n(\bm r_2,\bm r_1)=\lim_{\varepsilon\to \varepsilon_n} (\varepsilon-
\varepsilon_n)G(\bm r_2,\bm r_1|\varepsilon).
\end{eqnarray}

The quantity $\sigma_{bf}$ coincides with the total cross section of the
so-called bound-free pair production when an electron is produced in a bound
state. In fact, due to the Pauli principle, there is no bound-free pair
production  on neutral  atoms. Nevertheless,  the term $\sigma_{bf}$ should be
kept   in the r.h.s. of (\ref{eq:Opt}).  In a Coulomb field, the total cross
section $\sigma_{bf}$  was obtained  in \citep{MS1993} for $\omega\gg m$. In
this limit, $\sigma_{bf}\propto 1/m\omega$ and should be taken into account
when using the relation (\ref{eq:Opt}) for the calculation of the corrections
to $\sigma_{coh}$ from. The main contribution to $\sigma_{bf}$ comes from the
low-lying bound states \citep{MS1993} when screening can be neglected. So, in
(\ref{eq:Opt}) we can use $\sigma_{bf}$ obtained in \citep{MS1993}.

It is convenient to represent $d\sigma_{coh}/d\varepsilon_-$ and $M_{D}$ in
another form using the Green function $D(\bm{r}_2,\bm{r}_1|\varepsilon)$ of the
squared Dirac equation,
\begin{equation}\label{eq:FGD}
G(\textbf{r}_2,\textbf{r}_1 |\varepsilon )= \left[ \gamma^{0}(\varepsilon-
V(\textbf{r}_2))-\bm{\gamma}\textbf{p}_2+m \right] D(\textbf{r}_2,\textbf{r}_1
|\varepsilon )\,,\quad \bm p_2=-i\bm\nabla_2
\end{equation}

According to \citep{LM1995}, we can rewrite Eq. (\ref{eq:se}) in the form
\begin{eqnarray}\label{eq:spectrD}
\frac{d\sigma_{coh}}{d\varepsilon_-}&=&\frac{\alpha}{2\omega}
 \int\!\!\!\!\int\!\! d\bm{r}_1 d\bm{r}_2\,\mbox{e}^{-i\bm k\bm r}\nonumber\\
&& \times\mbox{Sp}\{
 [(2\bm e\bm p_2-\hat e\hat k)\delta D(\bm r_2 ,\bm r_1 |\varepsilon_-)]
 [(2\bm e\bm p_1+\hat e\hat k) \delta D(\bm r_1 ,\bm r_2|-\varepsilon_+)]\}
 \,,
\end{eqnarray}
and Eq. (\ref{eq:Md}) as
\begin{eqnarray}\label{eq:MdD}
M_D&=&i\alpha\int\!\!
d\varepsilon\int\!\!\!\!\int\!\! d\bm{r}_1 d\bm{r}_2\,\mbox{e}^{-i\bm k\bm r}\nonumber \\
&& \times\mbox{Sp}\{
 [(2\bm e\bm p_2-\hat e\hat k) D(\bm r_2,\bm r_1|\omega -\varepsilon)]
 [(2\bm e\bm p_1+\hat e\hat k) D(\bm r_1,\bm r_2 |-\varepsilon)]\}
\nonumber\\
&& +2i\alpha\, \int\!\! d\varepsilon\int\!\! d\bm{r}\, \mbox{Sp} D(\bm r,\bm r
|\varepsilon) \,.
\end{eqnarray}

The last term in (\ref{eq:MdD}) is $\omega$-independent, and has no imaginary
part. Therefore, it does not contribute to the relation (\ref{eq:Opt}).

\section{Green function}

To obtain the spectrum (\ref{eq:se}), (\ref{eq:spectrD}) and the Delbr\"uck
scattering amplitude (\ref{eq:Md}), (\ref{eq:MdD}) it is necessary to know the
explicit form of the Green function of the Dirac equation in the Coulomb
potential $V(r)=-Z\alpha/r$. An integral representation for
$G(\textbf{r}_2,\textbf{r}_1 |\varepsilon )$ has been obtained in
\citep{MS1982}. For $|\varepsilon|>m$ it has the form
\begin{eqnarray}\label{eq:Gexact}
&&G(\bm r_2,\bm r_1|\varepsilon)=-\frac i{4\pi r_2r_1 \kappa}\int_0^\infty ds
\exp[2iZ\alpha s\,\lambda+i\kappa (r_2+r_1) \coth s]\ T\nonumber\\
&&T=[1-(\bm\gamma\cdot\bm n_2)(\bm\gamma\cdot\bm
n_1)][(\gamma^0\varepsilon+m)\frac y2\, \partial_y S_B-iZ\alpha\gamma^0\kappa
\coth s \,S_B]\nonumber\\
&&+[1+(\bm\gamma\cdot\bm n_2)(\bm\gamma\cdot\bm n_1)](\gamma^0\varepsilon+m)
S_A +imZ\alpha\gamma^0\bm\gamma\cdot(\bm n_2+\bm n_1) S_B\nonumber\\
&&+\frac{i\kappa^2(r_2-r_1)}{2\sinh^2 s}\bm\gamma\cdot(\bm n_2+\bm n_1) S_B
-\kappa\coth s\ \bm\gamma\cdot(\bm n_2-\bm n_1) S_A\,.
\end{eqnarray}
In this formula
\begin{eqnarray}\label{eq:SaSb}
&&S_A=\sum_{l=1}^\infty \mbox{e}^{-i\pi\nu}J_{2\nu}(y)\,l
[P'_l(x)+P'_{l-1}(x)]\,,\quad
S_B=\sum_{l=1}^\infty\mbox{e}^{-i\pi\nu}J_{2\nu}(y)[P'_l(x)-P'_{l-1}(x)]\,,\nonumber\\
&&\nu=\sqrt{l^2-(Z\alpha)^2}\,,\quad \kappa=\sqrt{\varepsilon^2-m^2}\,,\quad
\lambda=\varepsilon/\kappa\,,\nonumber\\
&&y={2\kappa\sqrt{r_1r_2}}/{\sinh s}\,,\quad x=\bm n_1\cdot\bm n_2\,,\quad \bm
n_{1,2}={\bm r_{1,2}}/{r_{1,2}}\,,
\end{eqnarray}
$J_{2\nu}(y)$ are Bessel functions and $P_l(x)$ are Legendre polynomials,
$P'_l(x)=\partial_xP_l(x)$. The Green function $D(\bm r_2,\bm r_1|\varepsilon)$
can be obtained from (\ref{eq:Gexact}) by keeping in $T$ the terms $\propto m$:
\begin{eqnarray}\label{eq:Dexact}
D(\bm r_2,\bm r_1|\varepsilon)&=&-\frac i{4\pi r_2r_1 \kappa}\int_0^\infty ds
\exp[2iZ\alpha s\,\lambda+i\kappa (r_2+r_1) \coth s]\nonumber\\
&&\times\biggl\{[1-(\bm\gamma\cdot\bm n_2)(\bm\gamma\cdot\bm n_1)]\frac y2\,
\partial_y
S_B+[1+(\bm\gamma\cdot\bm n_2)(\bm\gamma\cdot\bm n_1)]S_A\nonumber\\
&& +iZ\alpha\gamma^0\bm\gamma\cdot(\bm n_2+\bm n_1) S_B\biggr\}\,.
\end{eqnarray}

We are going to derive the high-energy asymptotic expansion of the spectrum in
the region $\varepsilon_{\pm}\gg m$.  For the first two terms of such an
expansion the main contribution to the integral in (\ref{eq:se}),
(\ref{eq:spectrD}) is given by the region $r=|\bm r_2-\bm r_1|\sim \omega/m^2$,
see \citep{MS1983,MS1983a}. Let us introduce the variable $\bm\rho$ as the
component of $\bm r_1$ (or $\bm r_2$) perpendicular to $\bm r$:
\begin{equation}
\bm\rho=\frac{\bm r\times[\bm r_1\times\bm r_2]}{r^2}
\end{equation}

As shown in \citep{MS1983a}, the main contribution to the Coulomb corrections
to the spectrum originates from the region $\rho\sim 1/m$ and $\theta,\psi\sim
m/\omega\ll 1$, where $\theta$ is the angle between the vectors $\bm r_2$ and
$-\bm r_1$, and $\psi$ is the angle between the vectors $\bm r$ and $\bm k$. In
this region we have $\theta\approx r\rho/r_1r_2$. The argument of the Legendre
polynomials in (\ref{eq:SaSb}) is $x=\bm n_1\cdot\bm n_2\approx-1+\theta^2/2$.
Besides, the term $\kappa(r_2+r_1)\sim \omega^2/m^2\gg 1$  in the exponents in
(\ref{eq:Gexact}), (\ref{eq:Dexact}) is large, and  the integral is determined
by large $s$. Then $\coth s\approx 1+2 \exp(-2s)$, and from (\ref{eq:Dexact})
we have $\exp(-2s)\sim 1/\kappa r \sim m^2/\omega^2$. The argument, $y$, of the
Bessel functions in $S_{A,B}$ can be estimated as $y\sim \kappa r/\sinh s\sim
\omega/m\gg 1$.

A simple method of the calculation of $S_{A,B}$ at $y\gg 1$,
$1+x\approx\theta^2/2\ll1$, and $y\theta\sim 1$ has been formulated in the
Appendix of \cite{MS1983a}. It turns out that the leading term and the first
correction are determined by values of $l\sim y\sim \omega/m$ in sums
$S_{A,B}$. This fact is in agreement with the evident estimate $l\sim
\varepsilon\rho\sim \omega/m\gg 1$. In the same way as in \cite{MS1983a} we
obtain for $S_{A,B}$ with the first correction taken into account
\begin{equation}
S_A=-\frac{y^2}8J_0(y\theta/2)\left[1+i\frac{\pi(Z\alpha)^2}y\right]\,,\quad
S_B=-\frac{y}{2\theta}J_1(y\theta/2)\left[1+i\frac{\pi(Z\alpha)^2}y\right]
\end{equation}
Let us pass in (\ref{eq:Dexact}) from the integration over $s$ to the
integration over $y$, see (\ref{eq:SaSb}). We have
\[
 \exp[2iZ\alpha \lambda s]\approx\left(\frac{4\kappa\sqrt{r_1r_2}}y\right)^{2iZ\alpha\lambda} \,,\quad
 \coth s\approx 1+\frac{y^2}{8\kappa^2r_1r_2}\,.
\]
Then we obtain
\begin{eqnarray}\label{eq:Dquasi}
D(\bm r_2,\bm r_1|\varepsilon)&=&\frac {i\mbox{e}^{i\kappa(r_1+r_2)}}{16\pi
\kappa r_2r_1}\int_0^\infty
y\,dy\left(\frac{4\kappa\sqrt{r_1r_2}}y\right)^{2iZ\alpha \lambda}
\exp\left[\frac{i(r_2+r_1)y^2}{8\kappa r_1r_2}\right]\nonumber\\
&&\times\biggl\{[1+i\pi(Z\alpha)^2/y]
 \left[
 J_0(y\theta/2)+2iZ\alpha \frac{J_1(y\theta/2)}{y\theta}
 \bm\alpha\cdot(\bm n_2+\bm n_1)
 \right]
\nonumber\\
&&+\pi(Z\alpha)^2  \, \frac{J_1(y\theta/2)}{y^2\theta}[\bm n_2\times\bm
n_1]\cdot\bm\Sigma\biggr\}\,,
\end{eqnarray}
where $\bm\alpha=\gamma^0\bm\gamma$, $\bm
\Sigma=(i/2)[\bm\gamma\times\bm\gamma]$.
 This expression is the quasiclassical Green function of the
squared Dirac equation with the first correction taken into account. The
leading term in this expression, as well as the corresponding expression for
$G(\bm r_2,\bm r_1|\varepsilon)$ has been derived in \cite{MS1983,MS1983a}. It
is convenient to rewrite (\ref{eq:Dquasi}) in another form using the relations
\begin{equation}\label{eq:JtoExp}
\int_0^\infty\!\!\! q\, dq J_0(q\theta) g(q)=\int \frac{d\bm q}{2\pi}
\mbox{e}^{i\bm q\cdot\bm\theta} g(q)\,, \quad
 \int_0^\infty q\, dq
J_1(q\theta) g(q)=-i\int\!\!\! \frac{d\bm q}{2\pi}\frac{(\bm q\cdot
\bm\theta)}{q\theta} \mbox{e}^{i\bm q\cdot\bm\theta} g(q)\,,
\end{equation}
where $g(q)$ is an arbitrary function, $\bm q$ and $\bm\theta$ are
two-dimensional vectors. In our case we direct the vector $\bm\theta$ along
$\bm\rho$ so that $\bm\theta=r\bm \rho/r_1r_2$. Using (\ref{eq:JtoExp}) we have
\begin{eqnarray}\label{eq:Dquasi1}
D(\bm r_2,\bm r_1|\varepsilon)&=&\frac {i\mbox{e}^{i\kappa(r_1+r_2)}}{8\pi
 \kappa r_2r_1}\int d\bm q\left(\frac{2\kappa\sqrt{r_1r_2}}q\right)^{2iZ\alpha
\lambda}
\exp\left[\frac{i(r_2+r_1)q^2}{2\kappa r_1r_2}+i\bm q\cdot\bm\theta\right]\nonumber\\
&&\times\biggl\{\left[1+\frac{i\pi(Z\alpha)^2}{2q}\right]
 \left[1+Z\alpha \frac{\bm \alpha\cdot\bm q}{q^2} \right]
-\frac{i\pi(Z\alpha)^2}{4q^3}(\bm r/r)\cdot [\bm q\times \bm \Sigma
]\biggr\}\,.
\end{eqnarray}
The leading term of this formula has been obtained in \cite{LMS1997,LMS1998}.
Let us integrate by parts the term containing $\bm\alpha$-matrix and make the
change of variable $\bm q\to \kappa(\bm q-\bm \rho)$. We obtain
\begin{eqnarray}\label{eq:Dquasiclassical}
&&D(\bm{r}_2,\bm{r}_1 |\varepsilon )=
 \frac{i\kappa\mbox{e}^{i\kappa r}}{8\pi^2r_1r_2}\int d\bm{q}
 \exp\left[i\,\frac{\kappa r q^2 }{2r_1r_2} \right]
 \left(\frac{2\sqrt{r_1r_2}}{|\bm{q}-\bm{\rho}|}\right)^{2i Z\alpha \lambda}
 \\
 &&\times\left\{
 \left(1+\frac{r}{2r_1r_2\lambda}\bm{\alpha}\cdot \bm{q}\right)
 \left(1+i \frac{\pi(Z\alpha)^2}{2\kappa|\bm q-\bm \rho|}\right)
 -\frac{\pi(Z\alpha)^2}{4\kappa^2}(\gamma^0/\lambda-\bm \gamma\cdot \bm r/r)
 \frac{\bm \gamma\cdot (\bm q-\bm \rho)}{|\bm q-\bm\rho|^3}
 \right\}\,.\nonumber
\end{eqnarray}
Note that in this formula and below we can set $\lambda=\mbox{sign}\,
\varepsilon$.

It is easy to check, that within our accuracy the contribution of the last term
in braces vanishes after taking the trace in (\ref{eq:spectrD}). Therefore,
this term can be omitted in the problem under consideration. The remaining
terms in (\ref{eq:Dquasiclassical}) can be represented in the form
\begin{eqnarray}
&&D(\bm{r}_2,\bm{r}_1 |\varepsilon )=\left[1+\frac{\bm\alpha\cdot(\bm p_1+\bm
p_2)}{2\varepsilon}\right]D^{(0)}(\bm{r}_2,\bm{r}_1 |\varepsilon)\,,
\label{eq:DviaD0}
\\
&& D^{(0)}(\bm{r}_2,\bm{r}_1 |\varepsilon )= \frac{i\kappa\mbox{e}^{i\kappa
r}}{8\pi^2r_1r_2}\int d\bm{q}
 \exp\left[i\,\frac{\kappa r q^2 }{2r_1r_2}\right]
 \left(\frac{2\sqrt{r_1r_2}}{|\bm{q}-\bm{\rho}|}\right)^{2i Z\alpha \lambda}
 \left(1+i \frac{\pi(Z\alpha)^2}{2\kappa|\bm q-\bm \rho|}\right)\,.
\label{eq:D0quasiclassical}
\end{eqnarray}
The function $D^{(0)}(\bm{r}_2,\bm{r}_1 |\varepsilon )$ is nothing but the
quasiclassical Green function of the Klein-Gordon equation in the Coulomb
field. The function $\delta D$ in (\ref{eq:spectrD}) is defined as $\delta
D=D-\tilde{D}$, where $\tilde{D}$ is obtained from (\ref{eq:DviaD0}) by the
replacement $D^{(0)}\to D^{(0)*}$.

\section{Coulomb corrections to the spectrum}
\label{sec:CCS}

In this section we consider the Coulomb corrections to the spectrum,
$d\sigma_{C}/dx$, for $\varepsilon_{\pm}\gg m$ taking into account terms of the
order $m/\varepsilon_{\pm}$. According to \cite{DBM1954}, the higher order
terms of the perturbation theory with respect to the external field (Coulomb
corrections) are not seriously modified by screening. However, this question
has not been studied quantitatively so far. The influence of screening on
Coulomb corrections is investigated in detail in Section \ref{sec:Scr}. In the
present Section we calculate $d\sigma_C/d\varepsilon_-$ in a pure Coulomb
field.

\begin{figure}
\centering\setlength{\unitlength}{0.1cm}
\begin{picture}(105,80)
 \put(56,0){\makebox(0,0)[t]{$x$}}
 \put(-6,30){\rotatebox[origin=c]{90}{$\sigma_0^{-1} d\sigma_C/dx$}}
\put(0,0){\includegraphics[width=100\unitlength,keepaspectratio=true]{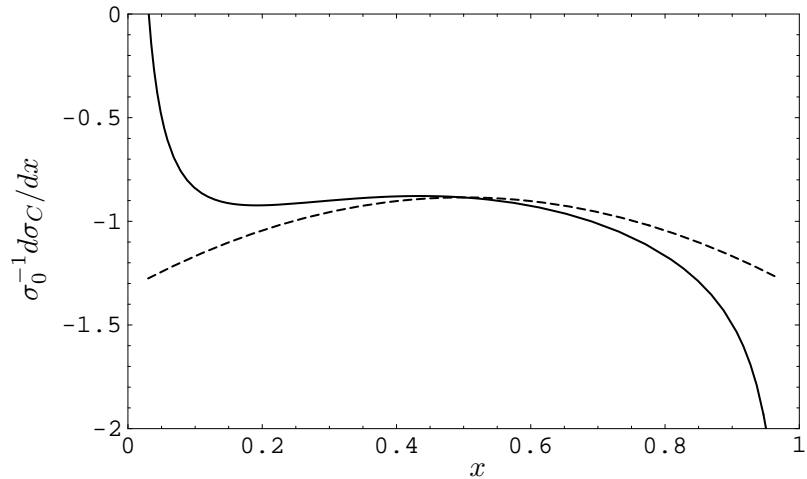}}
\end{picture}
\caption{The dependence of $\sigma_0^{-1} d\sigma_C/dx$ on $x$, see
(\ref{eq:spectr}), for $Z=82, \ \omega=50MeV$. Dashed curve: leading
approximation; solid curve: first correction is taken into account. }
\label{fig1}
\end{figure}

Substituting (\ref{eq:DviaD0}) in (\ref{eq:spectrD}) and taking the trace, we
obtain
\begin{eqnarray}\label{eq:D0-D0+}
&&\frac{d\sigma_{C}}{d\varepsilon_-}=
\frac{4\alpha}\omega\mbox{Re}\int\!\!\!\!\int d\bm r_1 d\bm r_2 \mbox{e}^{-i\bm
k\cdot \bm r}\biggl\{4 [\bm e\cdot\bm p_2 D^{(0)}_{-}][\bm e\cdot\bm p_1
D^{(0)}_{+}]\nonumber\\
&& -\frac{\omega^2}{\varepsilon_-\varepsilon_+}[\bm e\cdot(\bm p_1+\bm p_2)
D^{(0)}_{-}][\bm e\cdot(\bm p_1+\bm p_2) D^{(0)}_{+}]\biggr\}\,,\nonumber\\
&&D^{(0)}_{-}= D^{(0)}(\bm r_2,\bm r_1|\varepsilon_-)\,,\quad D^{(0)}_{+}=
D^{(0)}(\bm r_1,\bm r_2|-\varepsilon_+)\,.
\end{eqnarray}
The terms $\propto D^{(0)} D^{(0)*}$ are omitted in this formula since they do
not contribute to the leading term and the correction we are interested in.
Besides, we have integrated by parts the terms containing second derivatives of
$D^{(0)}$. In this formula and below we assume the subtraction from the
integrand the terms of the order $(Z\alpha)^0$ and  $(Z\alpha)^2$. Then we use
the relation
\begin{eqnarray}
(\bm e\cdot\bm p_{1,2})D^{(0)}(\bm{r}_2,\bm{r}_1 |\varepsilon )&=&
\frac{i\kappa^2\mbox{e}^{i\kappa r}}{8\pi^2r_1r_2}\int d\bm{q}
 \exp\left[i\,\frac{\kappa r q^2 }{2r_1r_2}\right]
 \left(\frac{2\sqrt{r_1r_2}}{|\bm{q}-\bm{\rho}|}\right)^{2i Z\alpha \lambda}
\nonumber\\
&&
 \times\left(1+i \frac{\pi(Z\alpha)^2}{2\kappa|\bm q-\bm \rho|}\right)
\left(\mp \frac{\bm e\cdot \bm r}r+\frac{\bm e\cdot\bm q}{r_{1,2}} \right)
 \,,
\label{eq:epD0}
\end{eqnarray}
and pass from the variables $\bm r_{1,2}$ to the variables
\begin{equation}
\bm r=\bm r_2-\bm r_1,\quad \bm\rho=\frac{\bm r\times[\bm r_1\times\bm
r_2]}{r^2},\quad z=-\frac{(\bm r\cdot\bm r_1)}{r^2}\, .
\end{equation}
In terms of these variables $d\bm r_1d\bm r_2=r\, d\bm r\, d\bm \rho\, dz$, and
within our accuracy $r_1=r z$, $r_2=r (1-z)$. We obtain from (\ref{eq:D0-D0+})
\begin{eqnarray}\label{eq:spectrInit}
&&\frac{d\sigma_{C}}{d\varepsilon_-}=
-\frac{\alpha\varepsilon_-\varepsilon_+}{16\pi^4\omega} \mbox{Re}\int
\frac{d\bm r}{r^5}\int_0^1 \frac{dz}{z^2(1-z)^2} \int\!\!\!\!\int\!\!\!\!\int
d\bm q_-d\bm
q_+ d\bm \rho\left(\frac{Q_+}{Q_-}\right)^{2iZ\alpha}\nonumber\\
&& \times\exp\left[\frac{i\omega r}2\left(
\psi^2-\frac{m^2}{\varepsilon_-\varepsilon_+}\right)
+i\frac{\varepsilon_-q_-^2+\varepsilon_+q_+^2}{2r
z(1-z)}\right]\left[1+\frac{i\pi(Z\alpha)^2}2
\left(\frac1{\varepsilon_-Q_-}+\frac1{\varepsilon_+Q_+}\right)\right]\nonumber\\
&& \times\left\{ 4\varepsilon_-\varepsilon_+ \left(\bm e\cdot\bm r+\frac{\bm
e\cdot\bm q_-}{1-z}\right)\left(-\bm e\cdot\bm r+\frac{\bm e\cdot\bm
q_+}z\right) -\frac{\omega^2}{z^2(1-z)^2}(\bm e\cdot \bm q_-)(\bm e\cdot \bm
q_+)
 \right\}\,,
\end{eqnarray}
where $Q_\pm=|\bm q_\pm-\bm\rho|$ and $\psi$ is the angle between vectors $\bm
r$ and $\bm k$. Since $d\sigma_C/d\varepsilon_-$ is independent of the photon
polarization, we can replace in (\ref{eq:spectrInit}) $e^ie^j$ by
$\frac12\delta^{ij}_\perp=\frac12(\delta^{ij}-k^ik^j/\omega^2)$. The integral
over $\bm\rho$ can be taken with the help of the relations (see Appendix)
\begin{eqnarray}\label{eq:fg}
&& f(Z\alpha)=\frac1{2\pi(Z\alpha)^2q^2} \int d\bm\rho
\left[\left(\frac{Q_+}{Q_-}\right)^{2iZ\alpha}\!\!\!-1+2(Z\alpha)^2
\ln^2\frac{Q_+}{Q_-}\right]=\mbox{Re}[\psi(1+iZ\alpha)+C]\nonumber\\
&&g(Z\alpha)=\frac i{4\pi q} \int \frac{d\bm\rho}{Q_+}
\left[\left(\frac{Q_+}{Q_-}\right)^{2iZ\alpha}\!\!\!-1\right] =
Z\alpha\,\frac{\Gamma(1-iZ\alpha)\Gamma(1/2 +i
Z\alpha)}{\Gamma(1+iZ\alpha)\Gamma(1/2 -i Z\alpha)}\,,
\end{eqnarray}
where $\psi(t)=d \ln \Gamma(t)/dt$, $C=0.577...$ is the Euler constant, $
q=|\bm q_--\bm q_+|$. We have
\begin{eqnarray}\label{eq:spectr1}
&&\frac{d\sigma_{C}}{d\varepsilon_-}=
-\frac{\alpha(Z\alpha)^2\varepsilon_-\varepsilon_+}{8\pi^2\omega}
\mbox{Re}\int_0^\infty\!\!\! \frac{dr}{r^3}\int_0^\infty\!\!\! \psi
d\psi\int_0^1 \frac{dz}{z^2(1-z)^2} \int\!\!\!\!\int d\bm q_-d\bm
q_+\nonumber\\
&& \times\exp\left[\frac{i\omega r}2\left(
\psi^2-\frac{m^2}{\varepsilon_-\varepsilon_+}\right)
+i\frac{\varepsilon_-q_-^2+\varepsilon_+q_+^2}{2r z(1-z)}\right] \left[ q^2
f(Z\alpha)+\pi q
\left(\frac{g(Z\alpha)}{\varepsilon_+}-\frac{g^*(Z\alpha)}{\varepsilon_-}
\right)\right]\nonumber\\
&& \times\left\{ 4\varepsilon_-\varepsilon_+ \left(-r^2\psi^2+\frac{\bm
q_-\cdot\bm q_+}{z(1-z)}\right) -\frac{\omega^2}{z^2(1-z)^2} (\bm q_-\cdot \bm
q_+)
 \right\}\,,
\end{eqnarray}
Passing to the variables $\tilde{\bm q}=\bm q_-+\bm q_+$, $\bm q=\bm q_--\bm
q_+$, we take all integrals in the following order: $d\psi$, $d\tilde{\bm q}$,
$d\bm q$, $dr$, $dz$. The final result for the Coulomb corrections to the
spectrum reads
\begin{eqnarray}\label{eq:spectr}
&&\frac{d\sigma_C^{(0)}}{dx}+\frac{d\sigma_C^{(1)}}{dx}=-{4\sigma_0}\biggl[
\left(1-\frac43 x(1-x) \right)
f(Z\alpha)\nonumber\\
&& -\frac{\pi^3(1-2x)m}{8x(1-x)\omega} \left(1-\frac32 x(1-x)\right)
\,\mbox{Re}\, g(Z\alpha) \biggr],\nonumber\\
&&x=\varepsilon_-/\omega\quad ,\quad \sigma_0=
\alpha(Z\alpha)^2/m^2\,.
\end{eqnarray}
In (\ref{eq:spectr}), the term $\propto
f(Z\alpha)$ corresponds to the leading approximation $d\sigma_C^{(0)}/dx$
\citep{DBM1954}, the term $\propto\mbox{Re}\, g(Z\alpha)$ is the first
correction $d\sigma_C^{(1)}/dx$. In contrast to the leading term, this
correction is antisymmetric with respect to the permutation
$\varepsilon_+\leftrightarrow\varepsilon_-$ (or $x\leftrightarrow 1-x$) and,
therefore, does not contribute to the total cross section. Besides, the
correction is an odd function of $Z\alpha$ due to the charge-parity
conservation and the antisymmetry mentioned above. The antisymmetric
contribution enhances the production of electrons  at $x<1/2$ and suppresses it
at $x>1/2$. Evidently, the opposite situation occurs for positrons.
Qualitatively, such a behavior of the spectrum  takes place for any $\omega$
being the most pronounced at low photon energy \citep{Overbo1968}. At
intermediate photon energies, the spectrum (\ref{eq:spectr}) essentially
differs from that given by the leading approximation. We illustrate this
statement in Fig.~\ref{fig1}, where $\sigma_0^{-1}d\sigma_C/dx$ with correction
(solid line) and without correction (dashed line) are plotted for $Z=82$ and
$\omega= 50\, MeV$.

Due to the antisymmetry of ${d\sigma_C^{(1)}}/{dx}$ at $\varepsilon_\pm\gg m$,
the term $\sigma_C^{(1)}$ in the total cross section may originate only from
the energy regions $\varepsilon_-\sim m$ and
$\varepsilon_+=\omega-\varepsilon_-\sim m$. The quasiclassical approximation
can not be used directly in these regions, and another approach is needed to
calculate the spectrum. We are going to do this elsewhere. However, for the
total cross section, it is possible to overcome this difficulty by means of
dispersion relations (see Section \ref{sec:CCT}).

As known \citep[see, e.g.,][]{BLP1980}, the spectrum of bremsstrahlung  can be
obtained from the spectrum of pair production. This can be performed by means
of the substitution $\varepsilon_+\to -\varepsilon$ , $\omega\to -\omega'$, and
$dx\to ydy$, where $y=\omega'/\varepsilon$, $\omega'$ is the energy of an
emitted photon, $\varepsilon$ is the initial electron energy. Using
(\ref{eq:spectr}), we obtain for the Coulomb corrections to the bremsstrahlung
spectrum
\begin{eqnarray}\label{eq:spectrphot}
y\frac{d\sigma_C^\gamma}{dy}&=&-{4\sigma_0}\biggl[
\left(y^2+\frac43(1-y) \right)
f(Z\alpha)\nonumber\\
&& -\frac{\pi^3(2-y)m}{8(1-y)\varepsilon} \left(y^2+\frac32
(1-y)\right) \,\mbox{Re}\, g(Z\alpha) \biggr]\, .
\end{eqnarray}

\begin{figure}
\centering\setlength{\unitlength}{0.1cm}
\begin{picture}(105,80)
 \put(56,0){\makebox(0,0)[t]{$x$}}
 \put(-6,30){\rotatebox[origin=c]{90}{$\sigma_0^{-1}y d\sigma^\gamma_C/dy$}}
\put(0,0){\includegraphics[width=100\unitlength,keepaspectratio=true]{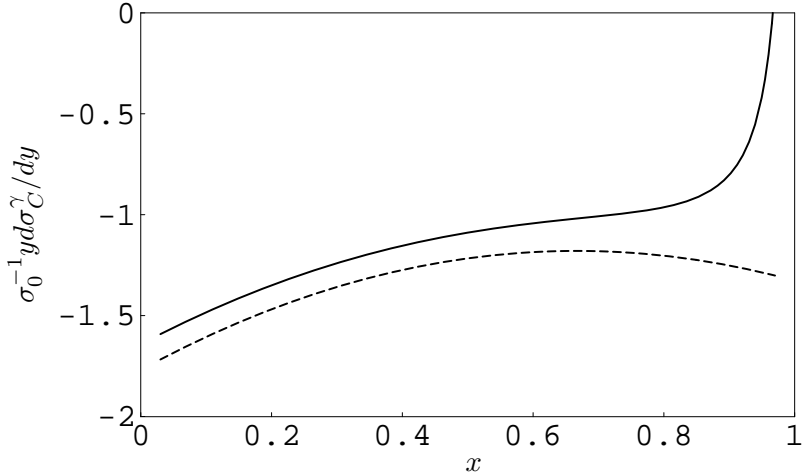}}
\end{picture}
\caption{The dependence of $\sigma_0^{-1} y d\sigma^\gamma_C/dy$ on $y$, see
(\ref{eq:spectr}), for $Z=82, \ \varepsilon=50MeV$. Dashed curve: leading
approximation; solid curve: first correction is taken into account. }
\label{fig1a}
\end{figure}

This formula describes bremsstrahlung from electrons. For the spectrum of
photons emitted by positrons, it is necessary to change the sign of $Z\alpha$
in (\ref{eq:spectrphot}). Our result (\ref{eq:spectrphot}) coincides  with that
obtained in \citep{BK1976} if the obvious mistake in the latter is corrected by
changing
$$\frac{1}{\gamma}\to \frac12\left(
\frac{m}{\varepsilon}+\frac{m}{\varepsilon-\omega'}\right)=
\frac{(2-y)m}{2(1-y)\varepsilon}\,
$$
in Eq.(22) of \citep{BK1976}. The correction (\ref{eq:spectrphot}) is the most
important at $y$ close to unity, see Fig. \ref{fig1a}, where $\sigma_0^{-1}y
d\sigma^\gamma_C/dy$ with correction (solid line) and without correction
(dashed line) are shown for $Z=82$ and $\varepsilon= 50\, MeV$.

\section{Coulomb corrections to the total cross section}
\label{sec:CCT}

 In the leading approximation, the Coulomb corrections, $\sigma_C^{(0)}$,
to the total cross section of pair production for $\omega\gg m$ were obtained
in \citep{DBM1954}. Using this result and dispersion relations, the
corresponding term, $M_{DC}^{(0)}$, in  the forward Delbr\"uck scattering
amplitude $M_D$ was obtained in \citep{Rohrlich1957}. These two quantities
read:
\begin{eqnarray}\label{eq:MD1}
\sigma_C^{(0)}=-\frac{28}{9}\,\sigma_0\,f(Z\alpha)\, ,\quad
M_{DC}^{(0)}=-i\frac{28}{9}\omega\sigma_0 \,f(Z\alpha) \, ,
\end{eqnarray}
where $\sigma_0$ and  $f(Z\alpha)$ are defined in (\ref{eq:spectr}) and
(\ref{eq:fg}), respectively.

In this Section we derive the correction $\sigma_C^{(1)}$ by means of the
relation (\ref{eq:Opt}). Starting from (\ref{eq:MdD}) and performing the same
calculations as in the previous Section we obtain
\begin{eqnarray}\label{eq:Md0Md1}
\label{eq:Mdspectr} &&M_{DC}^{(0)}+M_{DC}^{(1)}=-{4i\omega\sigma_0}\int_0^1 dx
\biggl[ \left(1-\frac43 x(1-x) \right)
f(Z\alpha)\nonumber\\
&& -\frac{\pi^3m}{8\omega} \left(1-\frac32 x(1-x)\right)
 \left(\frac{g^*(Z\alpha)}{x}-\frac{g(Z\alpha)}{1-x}\right) \biggr]\,.
\end{eqnarray}
Here the integration over $x$ corresponds to the integration over
$\varepsilon/\omega$. After the integration the $M_{DC}^{(0)}$ in
(\ref{eq:Md0Md1}) coincides with that in (\ref{eq:MD1}). The integral in
$M_{DC}^{(1)}$ is logarithmically divergent. Note that we have obtained the
integrand in (\ref{eq:MD1}) under the conditions $x\gg m/\omega$ and $1-x\gg
m/\omega$. Taking the integral from $\delta$ to $1-\delta$, where
$\delta\gtrsim m/\omega$, we find within logarithmic accuracy that $\mbox{Im}\,
M_{DC}^{(1)}$ vanishes and
\begin{equation}\label{eq:delM}  
\mbox{Re}\, M_{DC}^{(1)}=\,
\frac{\alpha(Z\alpha)^2\pi^3\,\mbox{Im}\,g(Z\alpha)}{m}\,\ln\frac{\omega}{m}
\,.
\end{equation}

The quantity $\mbox{Im}\, M_{DC}^{(1)}$ does not contain $\ln(\omega/m)$ and is
determined by the regions of integration over $\varepsilon$, where
$\varepsilon\sim m$ and $\omega-\varepsilon\sim m$, and, therefore, the
quasiclassical approximation is invalid. Nevertheless, this quantity, which is
related to $\sigma_C^{(1)}$ (\ref{eq:Opt}), can be obtained from the dispersion
relation for $M_D$ \citep{Rohrlich1957}
\begin{eqnarray}\label{eq:disp}
\mbox{Re}M_D(\omega)&=&\frac{2}{\pi}\omega^2\, P\!\int_0^\infty
\frac{\mbox{Im}M_D(\omega^{\prime}) \,d\omega'} {\omega^{\prime}(\omega^{\prime
2}-\omega^2)}\,.
\end{eqnarray}

Using this relation, it can be easily checked that the high-energy asymptotics
(\ref{eq:delM}) unambiguously corresponds to the $\omega$-independent
high-energy asymptotics
\begin{equation}\label{eq:delM1}  
 \mbox{Im}\, M_{DC}^{(1)}=
-\frac{\alpha(Z\alpha)^2\pi^4\,\mbox{Im}\,g(Z\alpha)}{2m} \, ,
\end{equation}

\begin{figure}[h]
\centering \setlength{\unitlength}{0.1cm}
\begin{picture}(105,80)
 \put(56,0){\makebox(0,0)[t]{$Z$}}
 \put(-6,35){\rotatebox[origin=c]{90}{$f_1$}}
 \put(0,0){\includegraphics[width=100\unitlength]{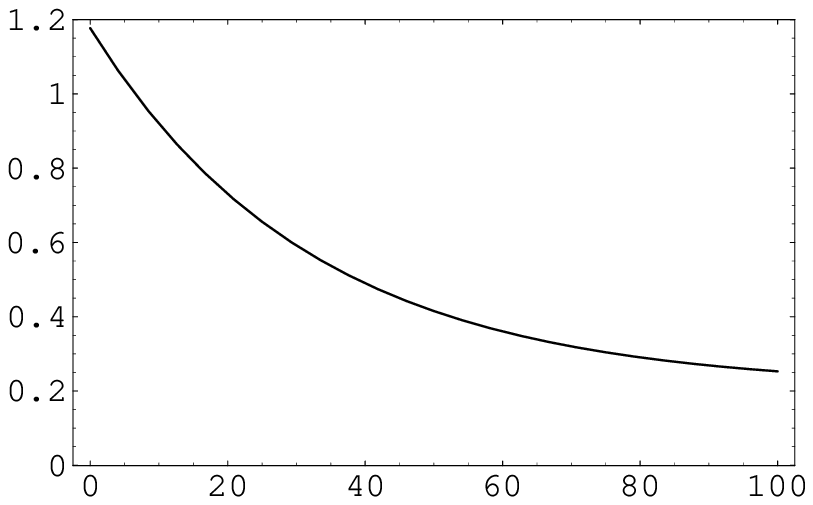}}
\end{picture}
\caption{The quantity $f_1$ as a function of $Z$}
 \label{fig2}\end{figure}

Substituting (\ref{eq:delM1}) into (\ref{eq:Opt}) and using $\sigma_{bf}$ from
\citep{MS1993} in the form
\begin{eqnarray}\label{eq:sigbf}
\sigma_{bf}&=&\,
4\pi\sigma_0\,(Z\alpha)^3\,f_1(Z\alpha)\,\frac{m}{\omega}\quad ,
\end{eqnarray}
we have for $\sigma_C^{(1)}$
\begin{eqnarray}\label{eq:delsigma} 
\sigma_C^{(1)}&=&-\sigma_0\,\left[\frac{\pi^4}{2}\, \mbox{Im}\,g(Z\alpha)
+4\pi(Z\alpha)^3\,f_1(Z\alpha)\right]\,\frac{m}{\omega}\quad .
\end{eqnarray}
The function $f_1(Z\alpha)$ is plotted in Fig.~\ref{fig2}.

\begin{figure}[h]
\centering \setlength{\unitlength}{0.1cm}
\begin{picture}(105,80)
 \put(56,0){\makebox(0,0)[t]{$Z$}}
 \put(-6,30){\rotatebox[origin=c]{90}{$(\omega/m)\sigma_C^{(1)}/\sigma_C^{(0)}$}}
\put(0,0){\includegraphics[width=100\unitlength]{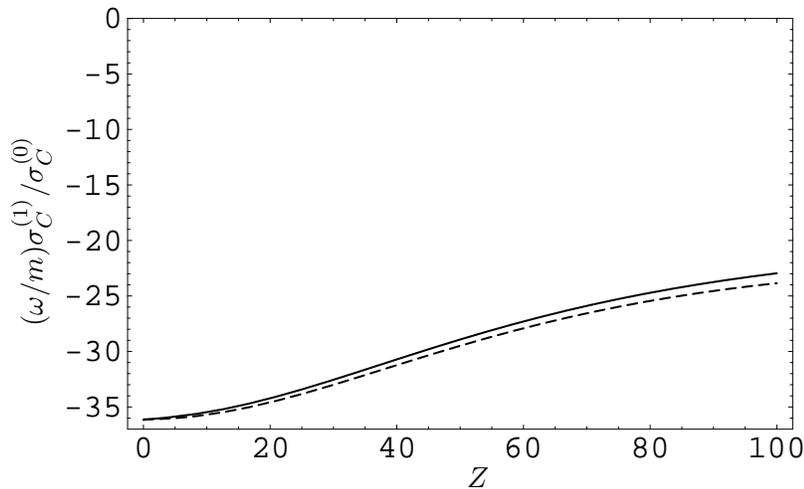}}
\end{picture}
\caption{The quantity $(\omega/m)\sigma_C^{(1)}/\sigma_C^{(0)}$ as a function
of $Z$(solid curve). Dashed curve corresponds to the same quantity without the
contribution of the bound-free pair production.}
 \label{fig2a}\end{figure}

The quantity $(\omega/m)\sigma_C^{(1)}/\sigma_C^{(0)}$ is shown in Fig.
\ref{fig2a} (solid curve). It is seen that this ratio is numerically large for
any $Z$. Therefore, the  term $\sigma_C^{(1)}$ gives a significant contribution
to $\sigma_C$ for intermediate photon energies. Dashed curve in Fig.
\ref{fig2a} gives the same ratio when $\sigma_{bf}$ in (\ref{eq:delsigma}) is
omitted. It is seen that the relative contribution of the term $\propto
f_1(Z\alpha)$ in (\ref{eq:delsigma}) is numerically small.

\section{Screening corrections}\label{sec:Scr}

In two previous Sections the cross section of $e^+e^-$ pair production has been
considered for a pure Coulomb field. The difference, $\delta V(r)$, between an
atomic potential and a Coulomb potential of a nucleus leads to the modification
of this cross section known as the effect of screening. In the Born
approximation, this effect was studied long ago \citep[see, e.g.,][]{JLS1950}.
Let us consider now $\sigma_C^{(scr)}$ characterizing the influence of
screening on the Coulomb corrections. Recollect that the Coulomb corrections
denote the higher-order terms of the perturbation theory with respect to the
atomic field. So far it was only known that the correction $\sigma_C^{(scr)}$
is not large \citep{DBM1954}. Here we consider this issue quantitatively.

The quasiclassical Green function $D^{(0)}(\bm r_2,\bm r_1|\varepsilon)$ for an
arbitrary localized potential $V(r)$ has been obtained in \cite{LMS2000} with
the first correction taken into account. The leading term has the form (see
also \cite{LM1995})
\begin{equation}\label{eq:D0V}
D^{(0)}(\bm r_2,\bm r_1 |\,\varepsilon )=\frac{i\kappa\mbox{e}^{i\kappa
r}}{8\pi^2r_1r_2} \int d\bm q \exp\left[i \frac{\kappa
r\,q^2}{2r_1r_2}-i\lambda r\int_0^1dx V\left(\bm r_1+x\bm r -\bm q\right)
\right] \,.
\end{equation}

Substituting this formula into (\ref{eq:D0-D0+}), we obtain (cf.
(\ref{eq:spectrInit}))
\begin{eqnarray}
\label{eq:spectrScrInit} &&\frac{d\sigma_{C}}{d\varepsilon_-}=
-\frac{\alpha\varepsilon_-\varepsilon_+}{16\pi^4\omega} \mbox{Re}\int
\frac{d\bm r}{r^5}\int_0^1 \frac{dz}{z^2(1-z)^2} \int\!\!\!\!\int\!\!\!\!\int
d\bm q_-d\bm q_+ d\bm \rho
\nonumber\\
&& \times\exp\biggl\{i\Phi+\frac{i\omega r}2\left(
\psi^2-\frac{m^2}{\varepsilon_-\varepsilon_+}\right)
+i\frac{\varepsilon_-q_-^2+\varepsilon_+q_+^2}{2r
z(1-z)}\biggr\}\nonumber\\
&& \times\left\{ 4\varepsilon_-\varepsilon_+ \left(\bm e\cdot\bm r+\frac{\bm
e\cdot\bm q_-}{1-z}\right)\left(-\bm e\cdot\bm r+\frac{\bm e\cdot\bm
q_+}z\right) -\frac{\omega^2}{z^2(1-z)^2}(\bm e\cdot \bm q_-)(\bm e\cdot \bm
q_+)
 \right\}\,,\nonumber\\
 &&
 \Phi=r \int_0^1 dx[V(\bm r_1+x\bm r -\bm q_+)-V(\bm r_1+x\bm r
-\bm q_-)]\,.
\end{eqnarray}
The phase $\Phi$ can be represented as
\begin{eqnarray}\label{eq:PhiScr}
\Phi&=&2Z\alpha \ln(Q_+/Q_-)+\Phi^{(scr)}\nonumber\\
&=&2Z\alpha \ln(Q_+/Q_-)+ r \int_0^1 dx[\delta V(\bm r_1+x\bm r -\bm
q_+)-\delta V(\bm r_1+x\bm r -\bm q_-)]\,.
\end{eqnarray}

As in the case of a pure Coulomb field, the main contribution to the Coulomb
corrections comes from the region of integration $q_\pm\sim\rho\sim 1/m$. The
main contribution to the integral over $x$ in (\ref{eq:PhiScr}) comes from the
narrow region around the point $x_0=-\bm r_1\cdot \bm r/r^2=z$, $\delta
x=\rho/r\ll 1$. Therefore, it is possible to perform the integration in
(\ref{eq:PhiScr}) from $-\infty$ to $\infty$. Thus we can estimate
$\Phi^{(scr)}$ as $\Phi^{(scr)}\sim \rho\, \delta V(\rho)\sim Z\alpha\, \delta
V(\rho)/ V(\rho)\ll 1$. In our calculation of $\sigma_C^{(scr)}$, we retain the
linear term of expansion in $\Phi^{(scr)}$. By definition,
\begin{equation}\label{eq:FF}
\delta V(r)=\int \frac{d\bm \Delta}{(2\pi)^3}\, e^{i\bm \Delta\bm
r}\,F(\Delta)\frac{4\pi Z\alpha}{\Delta^2}\, ,
\end{equation}
\begin{figure}[b]
\centering \setlength{\unitlength}{0.1cm}
\begin{picture}(105,80)
 \put(56,0){\makebox(0,0)[t]{$Z$}}
 \put(-6,30){\rotatebox[origin=c]{90}{$\sigma_C^{(scr)}/\sigma_C^{(0)}$}}
\put(0,0){\includegraphics[width=100\unitlength,keepaspectratio=true]{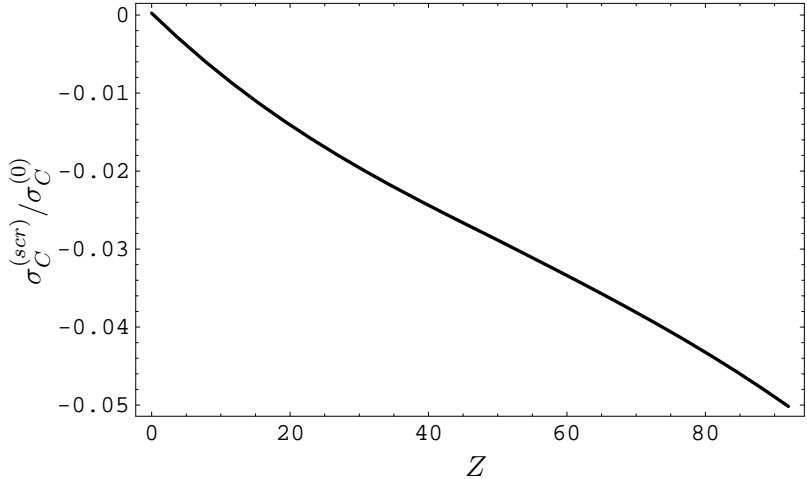}}
\end{picture}
\caption{The ratio $\sigma_C^{(scr)}/\sigma_C^{(0)}$ as a function of Z}
\label{fig3}\end{figure} where $F(Q)$ is the atomic electron form factor.
Substituting this formula to (\ref{eq:PhiScr}) and taking the integral over $x$
from $-\infty$ to $\infty$, we obtain for $\Phi^{(scr)}$
\begin{equation}  
\Phi^{(scr)}=\int \frac{d\bm \Delta_\perp}{(2\pi)^2}\, \left( e^{i\bm
\Delta_\perp(\bm \rho-\bm q_+)}-e^{i\bm \Delta_\perp(\bm \rho-\bm
q_-)}\right)\,F(\Delta_\perp)\frac{4\pi Z\alpha}{\Delta_\perp^2}\, ,
\end{equation}
where $\bm\Delta_\perp$ is two-dimensional vector lying in the plane
perpendicular to $\bm r$. Then we use the identity, see Eqs. (22), (23) in
\cite{LMS1998}
\begin{eqnarray}\label{eq:rho2f}
&&\int d\bm \rho \left(\frac{|\bm\rho-\bm q_+|}{|\bm\rho-\bm
q_-|}\right)^{2iZ\alpha}\exp\left[{i \bm\Delta_\perp\cdot (\bm \rho-\bm
q_\pm)}\right]\nonumber\\
&&= \frac{q^2}{4\Delta_\perp^2}\int d\bm f
\left(\frac{f_+}{f_-}\right)^{2iZ\alpha}\exp\left[{i \bm q\cdot \bm
f_\pm/2}\right]\,,
\end{eqnarray}
where $\bm q=\bm q_--\bm q_+$ and $\bm f_\pm=\bm f\pm\bm\Delta_\perp$.
Expanding (\ref{eq:spectrScrInit}) in $\Phi^{(scr)}$ and using the identity
(\ref{eq:rho2f}), we take the integrals over variables $\bm q_\pm$, $\bm r$,
and $z$ and obtain
\begin{eqnarray}\label{eq:scrspect1}
&&\frac{d\sigma_C^{(scr)}}{dx}=\frac{8\alpha(Z\alpha)}{3\pi} \int \frac{d\bm
\Delta_\perp}{\Delta_\perp^4}F(\Delta_\perp)
 \int\frac{d\bm f}{2\pi}\left(\frac{f_+}{f_-}\right)^{2iZ\alpha}
\left[\frac{R(\xi_-,a)}{f_-^2}-\frac{R(\xi_+,a)}{f_+^2} \right]\,,\nonumber\\
 &&R(\mu,\,a)=\frac{(\mu-1)}{4\mu^{2}}\Biggl\{
 \frac{1}{2\sqrt{\mu}}\left[18-6\mu+a(\mu^2+2\mu-3)\right]\,
 \ln\left[\frac{\sqrt{\mu}+1}{\sqrt{\mu}-1}\right]\nonumber\\
 &&-18-a(\mu-3)\Biggr\}\,,\nonumber\\
 &&\xi_\pm=1+16m^2/f_\pm^2\,
 ,\quad a=6x(1-x)\,.
\end{eqnarray}

Using the trick introduced in  Section IX in \cite{LMS1998}, we rewrite this
formula in another form. Let us multiply the integrand in (\ref{eq:scrspect1})
by
\begin{eqnarray}
\label{eq:ytrick} 1&\equiv&\int_{-1}^{1}dy\,\delta \left(y-\frac{2\bm f\cdot\bm
\Delta_\perp}{\bm f^2+\bm \Delta_\perp^2} \right)
\nonumber\\
&=&(\bm f^2+\bm \Delta_\perp^2)\int_{-1}^{1}\frac{dy}{|y|} \delta((\bm f-\bm
\Delta_\perp /y)^2 -\bm \Delta_\perp^2(1/y^2-1)) \,,
\end{eqnarray}
change the order of integration over $\bm f$ and $y$, and make the shift $\bm
f\rightarrow \bm f+\bm \Delta_\perp/y$. After that the integration over $f$ can
be done easily. Then we make the substitution $y=\tanh \tau$ and obtain
\begin{eqnarray}\label{eq:scrspect}
&&\frac{d\sigma_C^{(scr)}}{dx}=\frac{32}{3}\sigma_0m^2 \int_0^\infty
\frac{dQ}{Q^3}F(Q)\int_0^\infty\frac{d\tau}{\sinh
\tau}\left[\frac{\sin(2Z\alpha\tau)}{2Z\alpha}-\tau\right]\nonumber\\
&&\times\int_0^{2\pi}\frac{d\varphi}{2\pi}
\left[e^{\tau}R(\mu_+,\,a)-e^{-\tau}
 R(\mu_-,\,a)\right] \, ,\nonumber\\
  && \mu_\pm=
 1+\frac{8m^2e^{\pm\tau}\sinh^2\tau}{Q^2(\cosh\tau+\cos\varphi)}\, .
\end{eqnarray}
Integrating over $x$, we have
\begin{eqnarray}\label{eq:scr}
&&\sigma_C^{(scr)}=\frac{32}{3}\sigma_0m^2 \int_0^\infty
\frac{dQ}{Q^3}F(Q)\int_0^\infty\frac{d\tau}{\sinh
\tau}\left[\frac{\sin(2Z\alpha\tau)}{2Z\alpha}-\tau\right]\nonumber\\
&&\times\int_0^{2\pi}\frac{d\varphi}{2\pi}
\left[e^{\tau}R(\mu_+,\,1)-e^{-\tau}
 R(\mu_-,\,1)\right] \, \, .
\end{eqnarray}
Similar to $\sigma_C^{(0)}$, this correction is $\omega$-independent. Shown in
Fig.~\ref{fig3} is the $Z$-dependence of the ratio
$\sigma_C^{(scr)}/\sigma_C^{(0)}$ calculated with the use of the form factors
taken from \citep{HO1979}. As seen from Fig.~\ref{fig3}, this ratio is
approximately fitted by the linear function, $\sigma_C^{(scr)}\approx-5.4\cdot
10^{-4}\cdot Z\sigma_C^{(0)}$.

The corresponding correction to the bremsstrahlung spectrum is obtained from
(\ref{eq:scrspect}) by means of the same substitutions as in Section
\ref{sec:CCS}. So that the quantity $y^{-1}d\sigma_C^{\gamma(scr)}/dy$ is given
by the right-hand side of (\ref{eq:scrspect}) if we set $a=6(y-1)/y^2$. 

\section{Estimation of $\sigma_C^{(2)}$ from experimental data}

The most detailed and accurate experimental data have been obtained just in the
region of intermediate photon energies. In this region, the first correction
$\sigma_C^{(1)}$, obtained above, becomes large, see Fig. \ref{fig2a} for
$(\omega/m)\sigma_C^{(1)}/\sigma_C^{(0)}$, and the next term $\sigma_C^{(2)}$
in the expansion (\ref{eq:expansion}) may be significant. Using the arguments
similar to those presented by \citet{DBM1954} the following ansatz for
$\sigma_C^{(2)}$ has been suggested in our recent paper \cite{LMS2003}
\begin{equation}\label{eq:ansatz}
\sigma_C^{(2)}=\sigma_0 \left[b\ln(\omega/2m)+c\,\right] \left(\frac
m\omega\right)^2\,,
\end{equation}
where $b$ and $c$ are some functions of $Z\alpha$. It was shown in
\cite{LMS2003} that experimental data for $\sigma_{coh}$ are well described by
the formula
\begin{equation}\label{eq:SigmaCohFinal}
\sigma_{coh}=\sigma_B+\sigma_C^{(0)}+\sigma_C^{(scr)}+\sigma_C^{(1)}+\sigma_C^{(2)}\,,
\end{equation}
where $\sigma_C^{(0)}$, $\sigma_C^{(scr)}$, and $\sigma_C^{(1)}$ are from Eqs.
(\ref{eq:MD1}), (\ref{eq:scr}), and (\ref{eq:delsigma}), respectively;
$\sigma_C^{(2)}$ is given by (\ref{eq:ansatz}) with
$b=-3.78(\omega/m)\sigma_0^{-1}\sigma_C^{(1)}$, $c=0$.

\begin{figure}[h]
\centering \setlength{\unitlength}{0.1cm}
\begin{picture}(105,80)
 \put(56,0){\makebox(0,0)[t]{$\omega$ $(MeV)$}}
 \put(-4,30){\rotatebox[origin=c]{90}{$S$}}
\put(0,0){\includegraphics[width=100\unitlength,keepaspectratio=true]{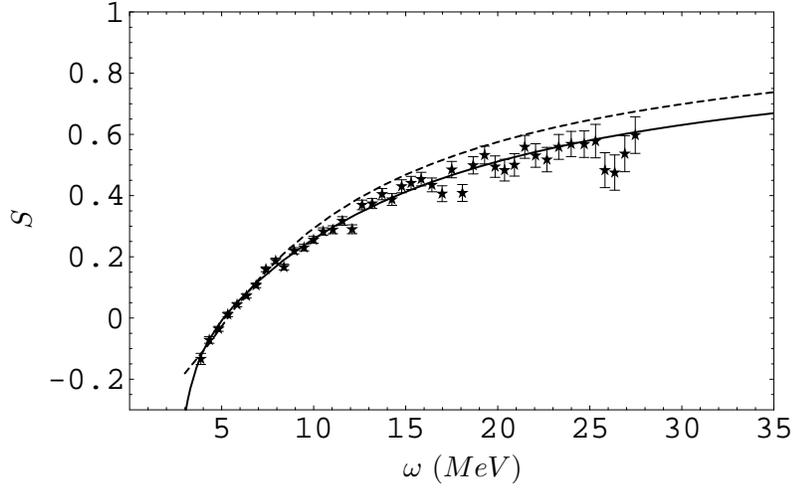}}
\end{picture}
\caption{The $\omega$-dependence of $S=(\sigma_{coh}-\sigma_B)/\sigma_C^{(0)}$
for $Bi$. Solid curve: our result; dashed curve: the result of
\citet{Overbo1977}; experimental data from \citep{SRL1980}.}
\label{fig7}\end{figure}

\begin{figure}[h]
\centering \setlength{\unitlength}{0.1cm}
\begin{picture}(105,80)
 \put(56,0){\makebox(0,0)[t]{$\omega$ $(MeV)$}}
 \put(-4,30){\rotatebox[origin=c]{90}{$S$}}
\put(0,0){\includegraphics[width=100\unitlength,keepaspectratio=true]{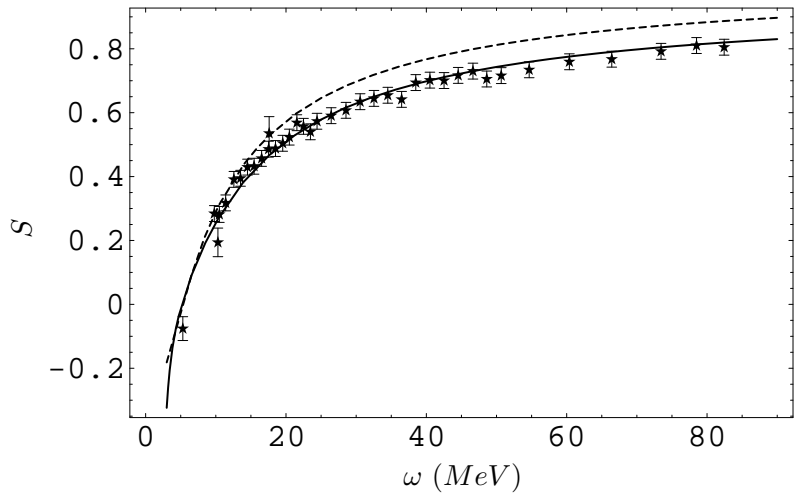}}
\end{picture}
\caption{Same as Fig. \ref{fig7} but for $Pb$; experimental data from
\citep{RSW1952,GH1978}.} \label{fig8}\end{figure}

It is interesting to compare our predictions for the Coulomb corrections to the
total cross section with the results of \citet{Overbo1977}. Shown in Figs.
\ref{fig7} and \ref{fig8} is the ratio
$S=(\sigma_{coh}-\sigma_B)/\sigma_C^{(0)}$, which is the Coulomb corrections in
units of $\sigma_C^{(0)}$, (\ref{eq:MD1}).
 Our results are represented
by solid curves, those of \citeauthor{Overbo1977} are shown as dashed curves.
The values of $S$ extracted from the experimental data are also shown. The
results for $Bi$ are plotted in Fig. \ref{fig7} with the experimental data
taken from \citep{SRL1980}.  The results for $Pb$ are plotted in Fig.
\ref{fig8} with the experimental data taken from \citep{RSW1952,GH1978}. It is
seen that the difference between our results and those of
\citeauthor{Overbo1977} is small at relatively low energies and becomes
noticeable as $\omega$ increases. According to our results, this difference
tends to a constant $\sigma_C^{(scr)}/\sigma_C^{(0)}$ at $\omega\to \infty$.
The experimental data are, on the whole, in a better agreement with our results
than with those of \citeauthor{Overbo1977}.

\section{Conclusion}

For the $e^+e^-$ photoproduction, we have calculated the leading correction
(\ref{eq:spectr}) to the electron spectrum in the region $\varepsilon_\pm\gg
m$. This contribution noticeably modifies the spectrum at intermediate photon
energy. It turns out that the correction is antisymmetric with respect to the
permutation $\varepsilon_+\leftrightarrow \varepsilon_-$ and hence does not
contribute to the total cross section. The leading correction to the total
cross section, $\sigma_C^{(1)}$, originates from two regions,
$\varepsilon_+\sim m$ and $\varepsilon_-\sim m$. We have obtained
$\sigma_C^{(1)}$ (\ref{eq:delsigma}) using dispersion relations. In contrast to
the form of the fit suggested by \citet{Overbo1977}, the quantity
$\sigma_C^{(1)}$ does not contain any powers of $\ln (\omega/ m)$. We have also
performed the quantitative investigation of the influence of screening on the
Coulomb corrections (\ref{eq:scrspect}),(\ref{eq:scr}). It is important that
$\sigma_C^{(scr)}$ does not vanish in the high-energy limit. We have suggested
a form for the next-to-leading correction, $\sigma_C^{(2)}$, to the total cross
section. Altogether, the corrections found allow one to represent well the
available experimental data.

Starting with the results obtained for  the  $e^+e^-$ photoproduction spectrum,
we have obtained the corresponding corrections to the bremsstrahlung spectrum
as well.

\begin{acknowledgments}
We are indebted to J.H.~Hubbell for his continuing interest to this work. This
work was supported in part by RFBR Grants 01-02-16926 and 03-02-16510.
\end{acknowledgments}

\appendix
\section*{Appendix}
\renewcommand{\theequation}{A\arabic{equation}}\setcounter{equation}{0}

In this appendix we derive the formulas (\ref{eq:fg}). In the integral for
$f(Z\alpha)$ let us make the change of variables $\bm\rho\to (\bm\rho+\bm
q_++\bm q_-)/2$:
\begin{equation}\label{eq:fInit}
f(Z\alpha)=\frac1{8\pi(Z\alpha)^2q^2} \int d\bm\rho
\left[\left(\frac{|\bm\rho+\bm q|}{|\bm\rho-\bm
q|}\right)^{2iZ\alpha}\!\!\!-1+2(Z\alpha)^2 \ln^2\frac{|\bm\rho+\bm
q|}{|\bm\rho-\bm q|}\right]
\end{equation}
Let us multiply the integrand in (\ref{eq:scrspect1}) by
\begin{eqnarray}
1&\equiv&\int_{-1}^{1}dy\,\delta \left(y-\frac{2\bm \rho\cdot\bm
q}{\rho^2+q^2} \right)
\nonumber\\
&=&(\bm \rho^2+\bm q^2)\int_{-1}^{1}\frac{dy}{|y|} \delta((\bm \rho-\bm q /y)^2
- q^2(1/y^2-1)) \,,
\end{eqnarray}
change the order of integration over $\bm \rho$ and $y$, and make the shift
$\bm \rho\rightarrow \bm \rho+\bm q/y$. Then the integral over $\bm \rho$
becomes trivial and we have
\begin{equation}
f(Z\alpha)=\frac1{4(Z\alpha)^2} \int_{-1}^1 \frac{dy}{|y|^3}
\left[\left(\frac{1+y}{1-y}\right)^{iZ\alpha}\!\!\!-1+\frac{(Z\alpha)^2}2
\ln^2\left(\frac{1+y}{1-y}\right)\right]
\end{equation}
Then we make the substitution $y=\tanh \tau$ and obtain
\begin{eqnarray}
f(Z\alpha)&=&\frac1{2(Z\alpha)^2} \int_{0}^\infty
d\tau\frac{\cosh\tau}{\sinh^3\tau}
\left[\cos(2Z\alpha\tau)-1+2(Z\alpha)^2\tau^2\right]\nonumber\\
&=&\int_{0}^\infty \frac{d\tau\mbox{e}^{-\tau}}{\sinh\tau}
\left[1-\cos(2Z\alpha\tau)\right]=\mbox{Re}[\psi(1+iZ\alpha)+C]\,.
\end{eqnarray}
In order to calculate the function $g(Z\alpha)$ we make a shift
$\bm\rho\to\bm\rho+\bm q_+$ in (\ref{eq:fg}) and use the exponential
parameterization
\begin{equation}
A^\nu=\frac{\mbox{e}^{i\pi\nu/2}}{\Gamma(-\nu)}\int_0^\infty\frac{ds}{s^{1+\nu}}
\exp[i A s]\,.
\end{equation}
Then we have
\begin{equation}
g(Z\alpha)=\frac {i\mbox{e}^{\pi Z\alpha/2}}{4\pi q \Gamma(iZ\alpha)} \int
d\bm\rho\, \rho^{-1+2iZ\alpha}\int_0^\infty ds\, s^{-1+iZ\alpha}
\left\{\exp[is(\bm\rho-\bm q)^2]-\exp[is\rho^2]\right\}\,.
\end{equation}
Taking the integral over the angles of two-dimensional vector $\bm\rho$ and
making the substitutions $\rho\to \rho/\sqrt{s}$, $s\to s/q^2$ we come to
\begin{equation}
g(Z\alpha)=\frac {i\mbox{e}^{\pi Z\alpha/2}}{2 \Gamma(iZ\alpha)} \int_0^\infty
d\rho\, \rho^{2iZ\alpha}\exp[i\rho^2]\, \int_0^\infty ds\, s^{-3/2}
\left[\mbox{e}^{is} J_0(2\rho\sqrt{s})-1\right]\,.
\end{equation}
Taking the integrals over $s$ and then over $\rho$ we finally obtain
\begin{equation}
g(Z\alpha)=Z\alpha\,\frac{\Gamma(1-iZ\alpha)\Gamma(1/2 +i
Z\alpha)}{\Gamma(1+iZ\alpha)\Gamma(1/2 -i Z\alpha)}\,.
\end{equation}

\end{document}